\documentclass[aps]{revtex4}
\usepackage[colorlinks=true, pdfstartview=FitV, linkcolor=blue, citecolor=red, urlcolor=magenta, breaklinks=true]{hyperref}
\usepackage{amsfonts}
\usepackage{graphics}
\usepackage{graphicx}
\begin{document}
\title{Lorentz violation correction to the Aharonov-Bohm scattering}
\author{M. A. Anacleto}
\email{anacleto@df.ufcg.edu.br}
\affiliation{Departamento de F\'{\i}sica, Universidade Federal de Campina Grande, Caixa Postal 10071,
58429-900, Campina Grande, Para\'{\i}ba, Brazil}

\begin{abstract}
In this paper, using a (2+1)-dimensional field theory approach, we study the Aharonov-Bohm (AB) scattering with Lorentz symmetry breaking. 
We obtain the modified scattering amplitude to the AB effect due to the small Lorentz violation correction   
in the breaking parameter and prove that up to one loop the model is free from ultraviolet divergences.  
\end{abstract}
\maketitle
\pretolerance10000
\section{Introduction}
The study of the Lorentz-symmetry-breaking theories suggested by Kostelecky {\it et al}.~\cite{Kost} has been intensively considered and constitutes  a fundamental tool in several fields.
The original motivation for this idea arose from the fact that 
the superstring theories suggest that Lorentz symmetry should be violated at higher energies.
In Ref.~\cite{Car}, Carrol {\it et al}. studied the Lorentz symmetry breaking in field theory. 
One of the interesting problems related to the Lorentz-symmetry-breaking QED was analyzed in Ref.~\cite{Nasc}. 
In Ref.~\cite{Bezerra2005} a spacetime with torsion interacting with a Maxwell field by means of a Chern-Simons-like term has been introduced. 
The authors in Ref.~\cite{Kanno2006} using a scalar-vector-tensor theory with Lorentz violation, 
have obtained an exact Lorentz-violation inflationary solution without an inflaton potential.
The Aharonov-Bohm-Casher problem  with a nonminimal  Lorentz-violating coupling was studied in Ref.~\cite{Belich2011}, and the authors have shown that the Lorentz violation is responsible
by the lifting of the original degeneracies in the absence of magnetic fields, even for a neutral particle.
Works have also been done on  topological defects with Lorentz symmetry violation~\cite{Dutra2006}. 
Investigations about monopoles and vortices due to Lorentz violation  were conducted in Ref.~\cite{Casana2012}.
Also, the problem of Lorentz-symmetry-violation  gauge theories in connection with gravity models was analyzed~\cite{Boldo2010}. 
In another work, Kostelecky and Mewes~\cite{Mewes2012} studied the effects of Lorentz violation 
in neutrinos~\cite{Almeida:2011ca}.
The authors in Ref. \cite{Fer} successfully realized the dimensional reduction of the Carrol-Field-Jackiw model to (2+1) dimensions. The study of some phenomenological implications of the three-dimensional ``mixed"
scalar-vector quadratic term has been analyzed~\cite{Gaete}.

In planar physics, the Aharonov-Bohm (AB) effect~\cite{Bohm} has been the object of several investigations. 
This effect is essentially the scattering of charged particles
by a flux tube and has been experimentally confirmed by Tonomura~\cite{RGC} (for review, see Ref.~\cite{Peskin}). 
In  quantum field theory the effect has been simulated, for instance, 
by using a nonrelativistic field theory describing bosonic particles
interacting through a Chern-Simons (CS) field~\cite{BL}.  It was also found to have analogues in several physical systems such as gravitation~\cite{FV}, fluid dynamics~\cite{CL}, optics~\cite{NNK} and Bose-Einstein condensates~\cite{LO} appearing in a vast literature.
The noncommutative AB effect has been studied in the context of quantum mechanics~\cite{FGLR,Chai} and in the quantum field theory approach~\cite{An, Anac}. 
In Ref.~\cite{FGLR}, it was shown that the cross section for the scattering of scalar particles by a
thin solenoid does not vanish even if the magnetic field assumes certain discrete values. 
In the context of quantum field theory, the effect was simulated, as in the commutative situation~\cite{BL}, 
by a nonrelativistic field theory of spin-0~\cite{An, Anac} and spin-$1/2$~\cite{An} particles
interacting through a  CS field.
The Aharonov-Bohm effect  for neutral particles based on the Lorentz-symmetry-violation background in the context of quantum mechanics was studied in Ref.~\cite{Bakke:2012gt}. 
However, a study of the AB effect with Lorentz symmetry breaking in the context of quantum field theory has not yet been realized.

Recently, it was shown in Ref.~\cite{Dolan} that the scattering of planar waves by a
draining bathtub vortex describes a modified AB effect which depends on two dimensionless parameters associated with the circulation and draining rates~\cite{Fetter}.  
In addition, we consider the acoustic black hole metrics obtained from a relativistic fluid
in a noncommutative spacetime~\cite{ABP12} via the Seiberg-Witten map and also obtained from the Lorentz-violating 
Abelian Higgs model~\cite{ABP11}. More recently in Ref.~\cite{ABP2012-1}, we have extended the analysis made in Ref.~\cite{Dolan}  to a Lorentz-violating and noncommutative background~\cite{Bazeia:2005tb} which allows us to have persistence of phase shifts even if circulation and draining vanish.

In this work we will further investigate the changes on the AB effect~\cite{Bohm} due to the Lorentz symmetry 
breaking in (2+1)-dimensional quantum field theory. We find a small Lorentz violation correction to the amplitude 
scattering.

\section{The Model}
The starting point of our studies is based on the (2+1)-dimensional model with Lorentz violation in the gauge sector described by the action
\begin{eqnarray}
\label{acao}
S[A,\phi]&=&\int d^3x \left\{\frac{\kappa}{2} \epsilon_{\mu\nu\lambda}A^{\mu}\partial^{\nu}A^{\lambda} 
+ \varphi\epsilon_{\mu\nu\lambda}v^{\mu}\partial^{\nu}A^{\lambda} 
- \frac{1}{2\xi}\partial_{i}A^{i}\partial_{j}A^{j}\right.
\nonumber\\
&&+\left.i\phi^{\dagger} D_{t}\phi -\frac{1}{2m}(D_{i}\phi)^{\dagger}(D_{i}\phi)
-\frac{\lambda_{0}}{4}\phi^{\dagger}\phi^{\dagger}\phi\phi
+ \frac{1}{2}\partial_{\mu}\varphi\partial^{\mu}\varphi\right\},
\end{eqnarray}
where the greek letters $ \mu $ run from $0$ to $2$ and $ i,j=1,2 $. 
The covariant derivatives in Eq.~(\ref{acao}) are given by
\begin{equation}
D_{t}\phi=\partial_{t}\phi + igA_{0}\phi +ig \varphi\phi,  \quad\quad\quad\quad  D_{i}\phi=\partial_{i}\phi + igA_{i}\phi.
\end{equation}
In the action we have considered the field $\phi $ interacting with the field $ \varphi $, and the inclusion of a quartic self-interaction of the scalar field $\phi$ (a nonrelativistic scalar field) is necessary to
secure the ultraviolet renormalizability of the model. 
The term $\varphi\epsilon_{\mu\nu\lambda}v^{\mu}\partial^{\nu}A^{\lambda}$, responsible for the Lorentz symmetry breaking,  was obtained by dimensional reduction in Ref.~\cite{Fer},
considering the following model in (3+1) dimensions:
 \begin{eqnarray}
 {\cal L}_{3+1}=-\frac{1}{4}F_{\hat{\mu}\hat{\nu}}F^{\hat{\mu}\hat{\nu}}
 +\frac{1}{2}\epsilon^{\hat{\mu}\hat{\nu}\hat{\rho}\hat{\lambda}}v_{\hat{\mu}}A_{\hat{\nu}}
 F_{\hat{\rho}\hat{\lambda}},
 \end{eqnarray}
where the greek letters $ \hat{\mu}=0,1,2,3$. The dimensional reduction is obtained by applying the following prescription to the gauge 4-vector, 
$ A^{\hat{\mu}} $, and
to the fixed external 4-vector, $ v^{\hat{\mu}} $:
\begin{eqnarray}
A^{\hat{\mu}} &\longrightarrow &(A^{\mu};\varphi) ,
\\
v^{\hat{\mu}} &\longrightarrow &(v^{\mu}; \kappa),
\end{eqnarray}
where $ A^{(3)}=\varphi $, $ v^{(3)}=\kappa $ and $ \mu=0,1,2 $. 
Thus the model in (2 + 1) dimensions is obtained:
\begin{eqnarray}
\label{lmred}
{\cal L}_{2+1}=-\frac{1}{4}F_{\mu\nu}F^{\mu\nu}
+\frac{\kappa}{2} \epsilon_{\mu\nu\lambda}A^{\mu}\partial^{\nu}A^{\lambda} 
+ \varphi\epsilon_{\mu\nu\lambda}v^{\mu}\partial^{\nu}A^{\lambda} 
- \frac{1}{2\xi}(\partial_{\mu}A^{\mu})^2+ \frac{1}{2}\partial_{\mu}\varphi\partial^{\mu}\varphi,
\end{eqnarray}
where the last term represents the gauge-fixing term, and that gives varying weight to the Lorentz gauge condition.
The $\varphi$ field also works out as the coupling constant in the term that mixes the gauge field 
to the fixed 3-vector, $v^\mu$. 
Further, the scalar field, $\varphi$, exhibits a typical Klein-Gordon massless dynamics.
As in quantum field theory, the Aharonov-Bohm effect is simulated by a nonrelativistic field theory describing bosonic particles interacting via a CS field, so for the model described by action (\ref{acao}), we consider
a scalar field nonrelativistic, $ \phi $, interacting with the fields $ A_0 $, $ A_i $ and $\varphi $, the Lagrangian (\ref{lmred}) without the Maxwell term.

Now neglecting divergence terms, the action (\ref{acao}) can be rewritten as 
\begin{eqnarray}
S[A,\phi]&=&\int d^3x\left\{A^{\mu}\left[\frac{\kappa}{2} \epsilon_{\mu\lambda\nu}\partial^{\lambda}
+\frac{1}{2\xi}\partial_{i}\partial_{j}\delta^{i}_{\mu}\delta^{j}_{\lambda}\right]A^{\nu} 
+\phi^{\dagger}\left[i\partial_{0}+\frac{\partial_i\partial_i}{2m}\right]\phi
+ \frac{1}{2}\varphi\left[\epsilon_{\mu\lambda\nu}v^{\mu}\partial^{\lambda}\right]A^{\nu}
+ \frac{1}{2} A^{\mu}\left[\epsilon_{\nu\lambda\mu}v^{\nu}\partial^{\lambda}\right]\varphi\right.
\nonumber\\
&-&\left.\frac{1}{2}\varphi\partial_{\mu}\partial^{\mu}\varphi-g\phi^{\dagger}A_0\phi
-g\phi^{\dagger}\varphi\phi
+\frac{ig}{2m}\left[\phi^{\dagger}A_j\partial_j\phi-(\partial_{j}\phi^{\dagger})A_j\phi \right]
-g^2\phi^{\dagger}A_jA_j\phi-\frac{\lambda_{0}}{4}\phi^{\dagger}\phi^{\dagger}\phi\phi\right\}.
\end{eqnarray}
Initially in our calculations, we shall choose $v^{\mu}$ to be purely timelike, $v^{\mu}=(v, {\vec 0})$, in the laboratory frame. 
Moreover, in our calculations for simplicity, we will work in a Coulomb
gauge obtained by letting $\xi\rightarrow 0$.
We will use a graphical notation where the CS field, the matter field $ \phi $ ,  the field $ \varphi $, and the mixed propagators are represented by wavy, continuous, dashed, and dashed-wavy lines, respectively.  

\begin{figure}
\centering
{\includegraphics{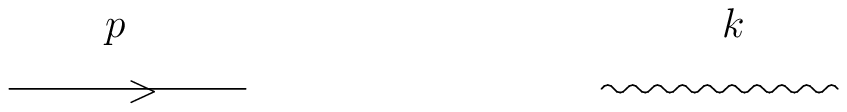}}
\caption{Propagators.}
\label{propagators}
\end{figure}

\begin{figure}
\centering
{\includegraphics[scale=.7] {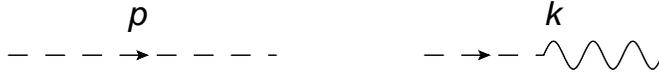}}
\caption{Field $ \varphi$ and mixed propagators.}
\label{propagators-mixed}
\end{figure}

\begin{figure}[htb]
\centering
{\includegraphics{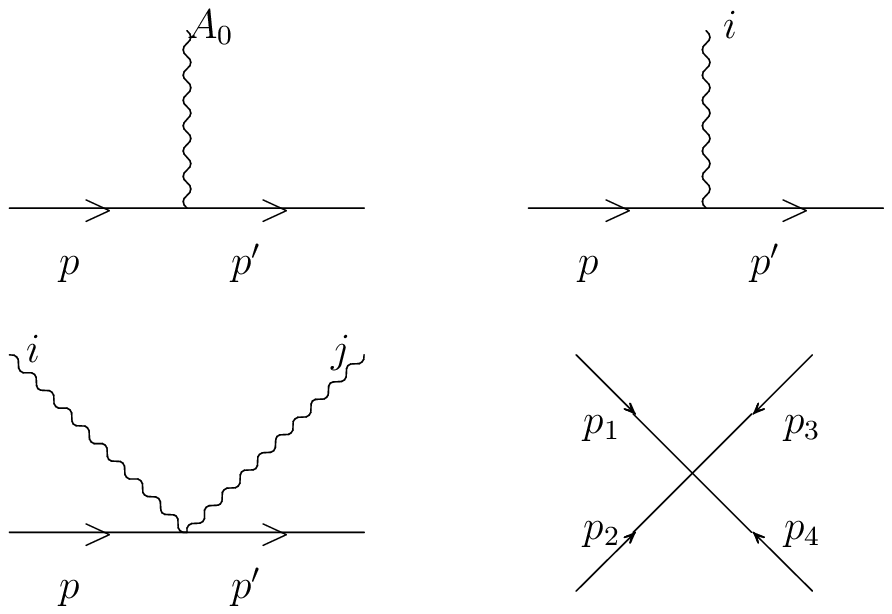}}
\caption{Vertices.}
\label{vertices}
\end{figure}

\begin{figure}[htb]
\centering
{\includegraphics[scale=.7] {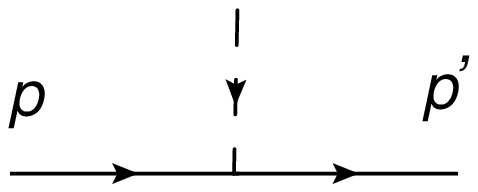}}
\caption{Vertices.}
\label{verticesvp}
\end{figure}

\begin{figure}
\centering
{\includegraphics{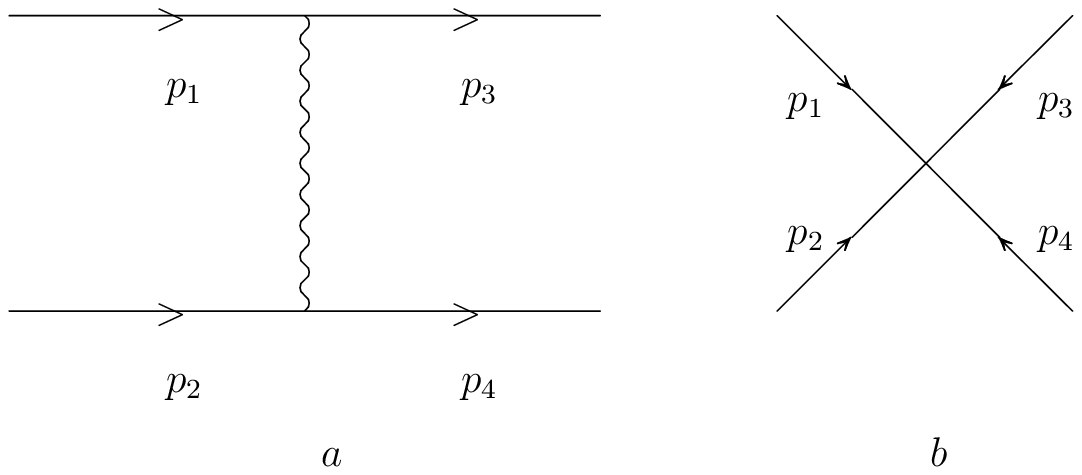}}
\caption{Tree level scattering.}
\label{treelevel}
\end{figure}

\begin{figure}[htb]
\centering
{\includegraphics[scale=.7] {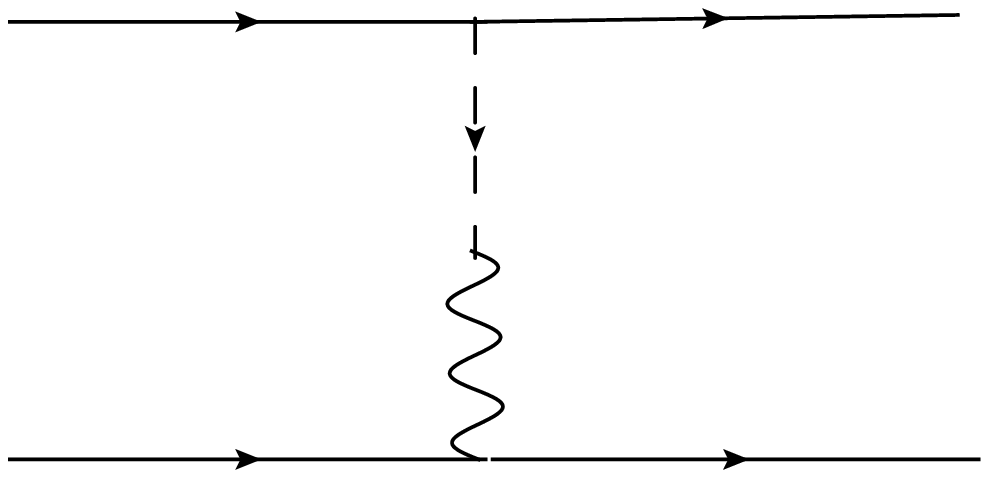}}
\caption{Tree level scattering.}
\label{tree-levelvp}
\end{figure}

The matter field and CS field propagators are (Fig.~\ref{propagators})
\begin{equation}
D(p)= \frac{i}{p_{0} -\frac{{\bf p}^2}{2m} + i\varepsilon},
\end{equation}
\begin{equation}
D_{i0}(k)=-D_{0i}(k)=\frac{\epsilon_{ij}{k}^{j}}{\kappa \bf {k}^2},
\end{equation}
\begin{equation}
\Delta(p)=\frac{i}{p^2},
\end{equation}
the mixed field propagators are (Fig.~\ref{propagators-mixed})
\begin{equation}
\left\langle A_{i}\varphi\right\rangle=-\left\langle\varphi A_{i}\right\rangle=-\frac{v\epsilon_{ji}k^{j}}{{\bf {k}}^4},
\end{equation}
and the vertices are given by (Figs.~\ref{vertices} and~\ref{verticesvp}) 
\begin{eqnarray}
&&\Gamma^{0}=-ig, \\
&&\Gamma_{\varphi} =-ig,\\
&&\Gamma^{i}=\frac{ig}{2m}{({p}+{p}^{\prime})}^{i}, \\
&&\Gamma^{ij}=-\frac{ig^2}{m}\delta ^{ij}, \\
&&\Gamma=-i\lambda_{0}.
\end{eqnarray}
At this point, we will realize a computation of the two-particle scattering at tree level in the center-of-mass frame.
Thus, for small $v$, we retain terms at first order in the parameter $v$. 
In the tree approximation, the two-body scattering amplitude is presented graphically 
in Figs.~\ref{treelevel} and~\ref{tree-levelvp}, corresponding to the following analytical expression: 
\begin{equation}
\label{amp1}
{\cal{A}}^{0}(\theta)=-\frac{2ig^2}{m\kappa}\cot\theta -\frac{16ig^{2}{\bar v}}{m}\frac{\cot^2\theta}{\sin(2\theta)}-\lambda_{0},
\end{equation}
where $\theta$ is the scattering angle between the incoming $({\bf p})$ 
and the outgoing $({\bf p}^{\prime})$ momenta, and ${\bar v}=\frac{v}{{\bf p}^2}$. 
Note that the second term in this amplitude~(\ref{amp1}) displays a small Lorentz violation correction 
in first order in the parameter $v$ and presents a different angular dependence of the result obtained 
in Refs.~\cite{An, Anac} in the noncommutative case.

\begin{figure}[htb]
\centering
{\includegraphics{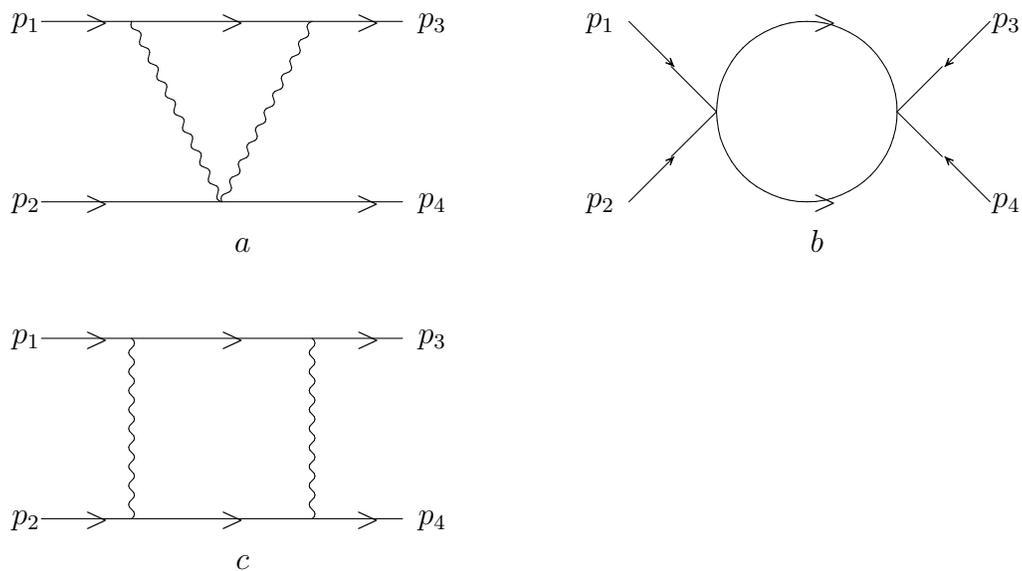}}
\caption{One-loop scattering diagrams.}
\label{loop}
\end{figure}

\begin{figure}
\centering
{\includegraphics[scale=.8] {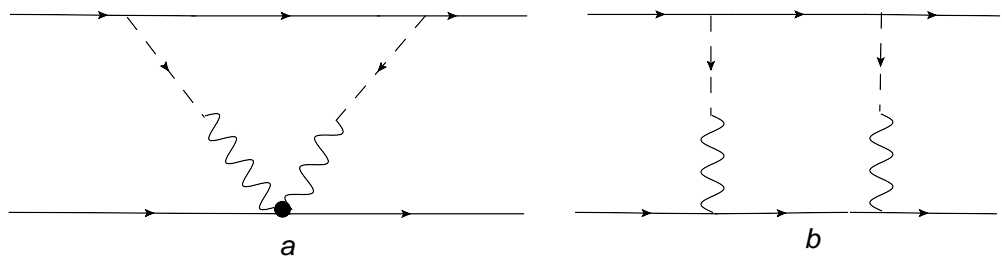}}
\caption{One loop scattering.}
\label{one-loopvp}
\end{figure}

The expressions for the contributions in one loop of the box and triangle graphs, are shown
in Figs. \ref{loop} and \ref{one-loopvp}. The other diagrams, correspoding to graphs [Fig.~\ref{loop}(a)]  with wavy lines exchanged by dashed-wavy lines, were not drawn. 
To compute the four-point function associated with the scattering of two identical particles, we separate their $v$-independent and $v$-dependent contributions:
\begin{eqnarray}
{\cal{A}}_{a}(\theta)&=&{\cal{A}}^{1}_{a}(\theta)+{\cal{A}}_{av}(\theta),
\\
{\cal{A}}_{b}(\theta)&=&{\cal{A}}^{1}_{b}(\theta)+{\cal{A}}_{bv}(\theta),
\\
{\cal{A}}_{c}(\theta)&=&{\cal{A}}^{1}_{c}(\theta)+{\cal{A}}_{cv}(\theta).
\end{eqnarray}
One should notice that ${\cal{A}}_{b}(\theta)$ does not present corrections in the parameter $v$, i.e., ${\cal{A}}_{bv}(\theta)=0$.

The calculations of the $v$-independent contributions are standard, so we just quote the results: 
after performing the $k_{0}$ integration,
for the triangle graph [Fig.~\ref{loop}(a)] we have~\cite{BL} 
\begin{eqnarray}
\label{graf-4a}
{\cal{A}}^{1}_{a}(\theta)&=&-\frac{g^4}{m\kappa^2}\int{\frac{d^2{\bf k}}{(2\pi)^2}
\frac{{\bf k}\cdot({\bf k}-{\bf q})}
{{\bf k}^2({\bf k}-{\bf q})^2}} + ({\bf p}_{3}\rightarrow -{\bf p}_{3}),
\nonumber\\
&=&\frac{g^4}{2\pi m\kappa^2}\left[\ln\left(\frac{\Lambda^2}
{{\bf p}^2}\right)-\ln(2\sin\theta)\right].
\end{eqnarray}
The result for the bubble diagram [Fig.~\ref{loop}(b)] reads
\begin{eqnarray}
{\cal{A}}_{b}(\theta)&=&\frac{m\lambda_0^2}{8\pi}\int_0^{\infty}d({\bf k}^2)
\frac{1}{{\bf k}^2-{\bf p}^2-i\epsilon},
\nonumber\\
&=&-\frac{m\lambda_{0}^{2}}{8\pi}\left[\ln\left(\frac{\Lambda^2}{{\bf p}^2}\right)+i\pi\right],
\end{eqnarray}
and that for the box graph [Fig.~\ref{loop}(c)]  is 
\begin{eqnarray}
\label{graf-4c}
{\cal{A}}^{1}_{c}(\theta)&=&\frac{4g^4}{m\kappa^2}\int_{0}^{\infty}{\frac{d^2{\bf k}}{(2\pi)}
\frac{({\bf p}_1\times {\bf k})\cdot ({\bf p}_3\times {\bf k})}
{({\bf k}-{\bf p}_1)^2({\bf k}-{\bf p}_3)^2({\bf k}^2-{\bf p}^2-i\varepsilon)}}+ ({\bf p}_{3}\rightarrow -{\bf p}_{3}),
\nonumber\\
&=&\frac{g^4}{2\pi m\kappa^2}[\ln(2\sin\theta)+i\pi],
\end{eqnarray}
where $\Lambda$ is an ultraviolet cutoff. 

Let us turn now to the computation of the $v$-dependent contributions. 
The lowest $v$-dependent correction to (\ref{graf-4a}) is given by Fig.~\ref{one-loopvp}(a);
after performing the $k_{0}$ integration, it is given as
\begin{eqnarray}
{\cal{A}}_{av}(\theta)&=&{\cal{A}}^{1}_{av}(\theta)+{\cal{A}}^{2}_{av}(\theta),
\end{eqnarray}
where
\begin{eqnarray}
\!\!\!\!{\cal{A}}^{1}_{av}(\theta)&=&-\frac{v^2g^4}{m}\int{\frac{d^2{\bf k}}{(2\pi)^2}
\frac{{\bf k}\cdot({\bf k}-{\bf q})}
{{\bf k}^4({\bf k}-{\bf q})^4}} + ({\bf p}_{3}\rightarrow -{\bf p}_{3}),
\\
\!\!\!\!{\cal{A}}^{2}_{av}(\theta)&=&\frac{-v^2g^4}{2m^2}\int{\frac{d^2{\bf k}}{(2\pi)^2}
\frac{({\bf p}_1\times{\bf k})({\bf p}_3\times{\bf k})+({\bf p}_1\times{\bf k})({\bf p}_{1}\times{\bf p}_{3})}
{{\bf k}^4({\bf k}-{\bf q})^4}}+ ({\bf p}_{3}\rightarrow -{\bf p}_{3}).
\end{eqnarray}
Here ${\bf q}={\bf p}_{1}-{\bf p}_{3}$ is the momentum transferred, and ${\bf k}\times{\bf p}\equiv \epsilon_{ij}k_ip_j$ is a ``vector" product of the two-dimensional spatial vectors which, however, in two-dimensional space is not a vector but a scalar. Both integrals can be evaluated analytically, i.e. 
\begin{eqnarray}
\label{av1}
\!\!\!\!{\cal{A}}^{1}_{av}(\theta)&=&-\frac{v^2g^4}{2m}\int_{0}^{\infty}{\frac{dk^2}{(2\pi)}
\frac{(k^2+{\bf q}^2)^2}
{k^2(k^2-{\bf q}^2)^3}} + ({\bf p}_{3}\rightarrow -{\bf p}_{3}),
\\
\!\!\!\!{\cal{A}}^{2}_{av}(\theta)&=&\frac{-v^2g^4}{4m^2}\int_{0}^{\infty}{\frac{dk^2}{(2\pi)}
\frac{[{\bf p}^2\cos\theta(k^2+{\bf q}^2)^2 + {\bf q}^2k^2(3{\bf p}^2-{\bf q}^2)]}
{k^2(k^2-{\bf q}^2)^4}}+ ({\bf p}_{3}\rightarrow -{\bf p}_{3}).
\label{av2}
\end{eqnarray}
Here $k^2\equiv\vec{k}^2$.
The $v$-dependent correction to (\ref{graf-4c}) is given by Fig.~\ref{one-loopvp}(b).
Following the same steps described for the previous case, we get
\begin{eqnarray}
\!\!\!\!\!\!\!\!\!{\cal{A}}_{cv}(\theta)&\!\!=\!\!&\frac{g^4v^2}{4m}\!\int_{0}^{\infty}\!\!{\frac{dk^2}{(2\pi)}
\frac{[8{\bf p}^2\cos\theta(k^2+{\bf p}^2)^{2}k^2 + k^2(2-4{\bf p}^2{\bf q}^2+{\bf q}^4)]}
{(k^2-{\bf p}^2)^8(k^2-{\bf p}^2-i\varepsilon)}}+ ({\bf p}_{3}\rightarrow -{\bf p}_{3}).
\label{cv}
\end{eqnarray}
It is interesting to note that the amplitudes (\ref{av1}), (\ref{av2}) and (\ref{cv}) are ultraviolet finite.

Thus, summing all the results in the one loop, we get
\begin{eqnarray}
\label{A1loop}
%{\cal{A}}(\theta)&=&{\cal{A}}_{a}(\theta)+{\cal{A}}_{b}(\theta)+{\cal{A}}_{c}(\theta),
%\nonumber\\
{\cal{A}}(\theta)&=&\frac{1}{2\pi m}\left(\frac{g^4}{\kappa^2}-\frac{m^2\lambda^2_0}{4}\right)
\left[\ln\left(\frac{\Lambda^2}{{\bf p}^2}\right)+i\pi \right]+ O(v^2).
\end{eqnarray}
For the special values, $\lambda_0=\pm\frac{2g^2}{m\kappa}$, the ultraviolet divergences vanish. 
Taking  this $ \lambda_0 $, we get the total amplitude one loop 
[= tree contribution (\ref{amp1}) + (\ref{A1loop})] in the form
\begin{eqnarray}
\label{ampt}
{\cal{A}}(\theta)&=&{\cal{A}}^{0}(\theta)+{\cal{A}}_{a}(\theta)+{\cal{A}}_{b}(\theta)+{\cal{A}}_{c}(\theta),
\nonumber\\
&=&-\frac{2ig^2}{m\kappa}\left[\cot\theta+8\kappa {\bar v}\frac{\cot^2\theta}{\sin(2\theta)}\right]\mp\frac{2g^2}{m\kappa} + O(v^2).
\end{eqnarray}
The Aharonov-Bohm scattering with Lorentz symmetry breaking is successfully obtained up to the one-loop order.
The choice of the lower or upper sign in (\ref{ampt}) corresponds to an attractive or repulsive quartic self-interaction. 
For a small angle $ \theta $, Eq. (\ref{ampt}) becomes
\begin{eqnarray}
\label{am-ap}
{\cal{A}}(\theta)=-\frac{2ig^2}{m\kappa}\left[\frac{1}{\theta}-\frac{\theta}{3}+O(\theta)^2
+8\kappa {\bar v}\left(\frac{1}{2\theta^3}-\frac{\theta}{30}+O(\theta)^3\right)\right]\mp\frac{2g^2}{m\kappa} + O(v^2).
\end{eqnarray}
Now the scattering amplitude at small angles, in the limit $ \theta\rightarrow 0 $, is dominated by
\begin{eqnarray}
\label{ampf}
{\cal{A}}(\theta)=-\frac{8ig^2}{m}\frac{\bar{v}}{\theta^3}.
\end{eqnarray}
Thus, the differential scattering cross section for small angles is
\begin{eqnarray}
\frac{d\sigma}{d\theta}=\vert {\cal A}(\theta)\vert^2\approx \frac{(8g^2)^2}{m^2}\frac{\bar{v}^2}{\theta^6}.
\end{eqnarray}
Thus, in the limit of $ \theta\rightarrow 0 $, the result for the differential cross section is due to only the contribution  
of $ {\bar v}^2 $ Lorentz symmetry breaking. This result is similar to that obtained in Ref.~\cite{ABP2012-1}.
On the other hand, considering $ v^{\mu} $ to be purely spacelike $ v^{\mu}=(0,{\bf v}) $, 
the mixed field propagators are 
\begin{eqnarray}
\left\langle\varphi A_0\right\rangle=-\left\langle A_{0}\varphi\right\rangle
=\frac{\epsilon_{ji}k^{j}v^{i}}{{\bf {k}}^4}, 
\\
\left\langle\varphi A_i\right\rangle=-\left\langle A_{i}\varphi\right\rangle
=\frac{\epsilon_{ij0}v^{j}k^{0}}{{\bf {k}}^4}.
\end{eqnarray}
In this case, the scattering amplitude in the tree approximation reads
\begin{equation}
\label{am-st}
{\cal{A}}^{0}(\theta)=-\frac{2ig^2}{m\kappa}\cot\theta 
 +ig^2\frac{\vert{\bf p}\vert\vert{\bf v}\vert }{4{\bf p}^4}
 \left[\frac{\sin\alpha-\sin\beta}{\sin^4(\theta/2)} +\frac{\sin\alpha+\sin\beta}{\cos^4(\theta/2)}\right]  -\lambda_{0},
\end{equation}
and that for a small $ \theta $ angle  becomes
\begin{eqnarray}
{\cal{A}}^{0}(\theta)&=&-\frac{2ig^2}{m\kappa}\left(\frac{1}{\theta}-\frac{\theta}{3}+O(\theta)^2 \right) 
%\nonumber\\
 +ig^2\frac{\vert{\bf p}\vert\vert{\bf v}\vert }{4{\bf p}^4}
 \left[\left(\sin\alpha-\sin\beta\right)\left(\frac{16}{\theta^4}+\frac{8}{3\theta^2}+\frac{11}{45}
 +\frac{31\theta^2}{1890}+O(\theta)^3 \right) \right]
\nonumber\\ 
&+&ig^2\frac{\vert{\bf p}\vert\vert{\bf v}\vert }{4{\bf p}^4}\left[\left(\sin\alpha+\sin\beta\right)
 \left( 1+\frac{\theta^2}{2} +O(\theta)^3\right)\right]  -\lambda_{0},
\end{eqnarray}
where $ \alpha $ is the angle
between $ {\bf p}_1 $ and ${\bf v} $,  and $ \beta $ is the angle between 
$ {\bf p}_3 $ and ${\bf v} $. Furthermore, the amplitude $v$ dependent, in the one loop order, does not present ultraviolet divergences. 
However, the AB amplitude in the limit $ \theta\rightarrow 0 $, is dominated by
\begin{eqnarray}
{\cal{A}}(\theta)=\frac{4ig^2\vert{\bf p}\vert\vert{\bf v}\vert }{{\bf p}^4\theta^4}\left(\sin\alpha-\sin\beta\right).
\end{eqnarray}
In this case, the differential scattering cross section for small angles becomes
\begin{eqnarray}
\frac{d\sigma}{d\theta}=\vert {\cal A}(\theta)\vert^2
\approx\frac{16g^4{\bf v}^2 }{{\bf p}^6\theta^8}\left(\sin\alpha-\sin\beta\right)^2.
\end{eqnarray}
%Note that the differential scattering cross section vanishes for both angles assuming the values $ \alpha,\beta=n\pi, (%n+1/2)\pi $, with $ n=0,1,2,\cdots $. 
Note that the differential scattering cross section vanishes 
for any angles satisfying $ \alpha=\beta $ and $ 0\le |\sin\alpha-\sin\beta|\le 1  $. 
Now, for example, if $ \alpha=0 $ and $ \beta=\pi/2 $, or $ \alpha=\pi/2 $ and $ \beta=0 $, we have
\begin{eqnarray}
\frac{d\sigma}{d\theta}\approx {(16g^4{\bf v}^2) }/{({\bf p}^6\theta^8)} .
\end{eqnarray} 
This correction vanishes in the limit $ {\bf v}^2\rightarrow 0 $ so that no singularities are generated. 
A contribution occurring in second order in the breaking parameter of the Lorentz symmetry to the cross section was also obtained in Ref.~\cite{ABP2012-1}.
This correction ($\sim {\bf v}^2$) due to the effect of Lorentz symmetry breaking may be relevant at high energies.

\section{Conclusion}
In this paper, we find that the scattering amplitude in the tree approximation displays a small Lorentz violation correction in first order in the parameter $v$ and contains an angular dependence. 
Moreover, we have found that the a  $v$-dependent amplitude, in the one-loop order, does not present ultraviolet divergences.
Also, we have shown that the correction to the amplitude in the one-loop order 
occurs only in the second order in the parameter $v$. 
The AB amplitude with Lorentz symmetry breaking in the limit $v\rightarrow 0$ agrees with the usual result~\cite{BL}. 
In addition, we show for each case, timelike and spacelike $ {\bf v} $, that the differential cross section at the small-angle limit is essentially due to the effect of the Lorentz symmetry breaking. Thus, in this limit the breaking of Lorentz symmetry strongly contributes to the AB effect.
However, our results allow an experimental verification of detecting Lorentz-symmetry-breaking signals 
via the Aharonov-Bohm effect.

\acknowledgments
The author would like to thank F. A. Brito and E. Passos for useful discussions and CNPq, CAPES, PNPD/PROCAD - CAPES for partial financial support. 
The author would also like to thank the referee for his valuable suggestion.

\newpage

\newpage

\end{document}